\documentclass[a4paper,11pt]{article}  
\usepackage{graphicx,amsmath,amssymb,dsfont,bm}
\usepackage{color}
\usepackage{enumitem}
\usepackage{boldline}
\usepackage{geometry}
\usepackage{authblk}
\usepackage[sort&compress, numbers]{natbib}
\usepackage{bibentry}
\usepackage[title]{appendix}
\usepackage{multirow}
\usepackage{makecell}
\usepackage{longtable}

\usepackage[normalem]{ulem}
\newcommand\redsout{\bgroup\markoverwith{\textcolor{red}{\rule[0.5ex]{2pt}{0.6pt}}}\ULon}

\title{\bf Local fluctuations of the signed traded volumes and the dependencies of demands: a copula analysis}

\author{Shanshan Wang\thanks{shanshan.wang@uni-due.de}~~and Thomas Guhr\thanks{thomas.guhr@uni-due.de}}
\affil{\textit{Fakult\"at f\"ur Physik, Universit\"at Duisburg--Essen, Lotharstra\ss e 1, 47048 Duisburg, Germany}}
\date{\today}
 
\begin{document}
\maketitle

\begin{abstract}
We investigate how the local fluctuations of the signed traded volumes affect the dependence of demands between stocks. We analyze the empirical dependence of demands using copulas and show that they are well described by a bivariate $\mathcal{K}$ copula density function. We find that large local fluctuations strongly increase the positive dependence but lower slightly the negative one in the copula density. This interesting feature is due to cross-correlations of volume imbalances between stocks. Also, we explore the asymmetries of tail dependencies of the copula density, which are moderate for the negative dependencies but strong for the positive ones. For the latter, we reveal that large local fluctuations of the signed traded volumes trigger stronger dependencies of demands than of supplies, probably indicating a bull market with persistent raising of prices.
\end{abstract}

\section{Introduction}

\label{sec1}

The demand drives the buying and selling of financial markets. It can be quantified by the traded volume imbalance, \textit{i.e.}, by the difference between the bought-in volumes and the sold-out volumes for an individual stock. In an influential study~\cite{Plerou2003}, Plerou, Gopikrishnan, and Stanley (PGS) discovered a two-phase behavior of demands in financial markets. The two-phase behavior shows itself in a transition from the unimodal distribution (one peak) of volume imbalances to a bimodal one (two peaks). It depends on the local noise intensity, defined as the absolute value of fluctuations around the average of volume imbalances in a certain time interval. Matia and Yamasaki (MY)~\cite{Matia2005} further investigated the causes of the two-phase behavior by estimating the volume imbalance and local noise intensity from Trades and Quotes data as well as from  a numerical simulation. They observed one or two phases, if the generated time series were equipped with Gaussian or power-law distributed traded volumes, respectively. The two-phase behavior may thus be interpreted as a consequence of fat-tailed distributions of traded volumes. This is similar to the conclusion proposed by Potters and Bouchard (PB)~\cite{Potters2003}, who argued that the high correlation between the noise intensity and the magnitude of the volume imbalance is the main reason for the symmetric distribution of the latter around zero. In other words, the two-phase behavior is due to the known statistical properties of traded volumes.

The interpretations of MY and PB are consistent with those of PGS~\cite{Plerou2005} who also elucidated the significance of the two-phase behavior in terms of the price change in corresponding market states. They demonstrated that the prices fluctuate slightly around the local equilibrium values when the traded volume is comparable with the market depth (corresponding to small demand or local noise intensity), whereas the price moves greatly when the traded volume is much larger than the depth (corresponding to large demand or local noise intensity)~\cite{Plerou2005}. Here, the large traded volume leads to a directional demand, \textit{i.e.} either buy or sell, as well as to a large local noise intensity. 

Other previous studies on the two-phase behavior also focus on its mechanism~\cite{Zheng2004,Sinha2004,Sinha2006,Lim2007}. Especially, some interpretations are put forward based on agent-based models~\cite{Sinha2004,Sinha2006}, minority games~\cite{Zheng2004} and herding models~\cite{Zheng2004}. In addition, the two-phase behavior is examined not only in the stock market, but also across future markets~\cite{Lim2007,Hwang2010}, option markets~\cite{Ryu2013} and financial indices~\cite{Jiang2008}. However, in all these studies only the statistical properties of individual stocks are taken into account. Here, we want to complete the picture by looking at the statistical dependence of demands across stocks and how this dependence structure depends on the local noise intensity. 

We employ copulas, introduced first by Sklar in 1959~\cite{Sklar1959,Sklar1973}. The idea behind the copula is to map all marginal distributions to uniform distributions and then to measure the joint distribution density as a function of the corresponding quantiles. Due to the separation between the pure statistical dependence of random variables and the marginal probability distributions, the copula has become an important, standard tool for directly modelling or comparing the statistical dependencies of different systems. Considering the importance of fat tails, we resort to a $\mathcal{K}$ copula density for the comparison with data. The $\mathcal{K}$ copula, introduced in references~\cite{Wollschlager2015,Chetalova2015b}, is based on a multivariate distribution in terms of a modified Bessel function of the second kind. This distribution results from a Random Matrix Model for the non-stationarity of financial data~\cite{Schmitt2013,Chetalova2015c} and was found to describe the empirical fat-tailed multivariate distributions of returns rather well~\cite{Schmitt2013,Schmitt2014,Chetalova2015c,Chetalova2015a}. Here, we will show that, in contrast to the Gaussian copula density, the $\mathcal{K}$ copula density also gives a good description of the empirical dependence of demands. We also provide a further view on the asymmetry of the tail dependencies of demands, and demonstrate the influence of the large local noise intensity on the dependence structure. 

The paper is organized as follows. In section~\ref{sec2}, we introduce the data set, give some basic definitions, and present the demand distributions of individual stocks. In section~\ref{sec3}, to analyze the empirical dependencies between stocks, we review the concept of the copula density, discuss the estimation method of the empirical copula density, and show the empirical results, including the empirical copula density between stocks and its tail asymmetries. In section~\ref{sec4}, we fit the empirical copula density by a bivariate $\mathcal{K}$ copula density function and a Gaussian copula density function, and compare the two fit results. In section~\ref{sec5}, we investigate the influences of local fluctuations on the dependence of demands and on the tail asymmetries of the dependencies. We conclude and discuss our results in section~\ref{sec6}.

\section{Data set, trade signs and demand distributions}
\label{sec2}
We present our data set in section~\ref{sec21}, and give basic definitions of trade signs in section~\ref{sec22}. In section~\ref{sec23}, we define the demands and examine the effect of the local noise intensity on the marginal distribution of demands. 

\subsection{Data set}
\label{sec21}
The stocks are from NASDAQ stock market in the year 2008, where all successive transactions and quotes of those stocks are recorded in Trades and Quotes (TAQ) data set. To avoid overnight effects and the drastic fluctuations at the opening and closing of the market, we exclude the trades occurring in the first and the last 10 minutes of the trading time for each day. For a stock pair, only the common trading days are taken into account for calculating the copula densities, because the dependence between stocks is absent when either stock does not trade. In section~\ref{sec23}, to calculate the conditional probability density distributions, we use 496 available stocks from S$\&$P 500 index in 2008. For the empirical copula densities to be evaluated in sections~\ref{sec3} and \ref{sec5}, we select the first 100 stocks, listed in \ref{appA}, with the largest average number of daily trades among those 496 stocks. The number of daily trades, also excluding the ones in the first and the last 10 minutes of the daily trading time, is averaged over a whole year for each stock.

\subsection{Trade signs}
\label{sec22}

In a time interval labeled $t$, various trades with running number $n$ may occur with corresponding prices $S(t;n)$. Each such trade in the TAQ data set can be classified as either buyer-initiated or seller-initiated~\cite{Wang2016a,Wang2016b} by
\begin{eqnarray}       
\varepsilon(t;n)=\left\{                  
\begin{array}{lll}    
\mathrm{sgn}\bigl(S(t;n)-S(t;n-1)\bigr)    &,& ~~\mbox{if}~~S(t;n)\neq S(t;n-1) , \\    
            \varepsilon(t;n-1) &,&~~\mbox{otherwise},
\end{array}           
\right.  
\label{eq2.2.1}            
\end{eqnarray}
where $\varepsilon(t;n)$ represents the sign of $n$-th trade in a time interval. A trade is identified as buyer-initiated if $\varepsilon(t;n)=1$, and a seller-initiated if $\varepsilon(t;n)=-1$. Zero values for $\varepsilon(t;n)$ are absent, because we do not aggregate the trade signs in a physical time interval as in our previous studies~\cite{Wang2016a,Wang2016b}. It is worth mentioning that due to the resolution of one second in the TAQ data set, the algorithm of Lee and Ready~\cite{Lee1991} cannot be used to classify the trades occurring in a time interval of one second. Instead, equation~(\ref{eq2.2.1}) is designed to classify continuous trades in smaller time scale than one second, too.

\subsection{Demand distributions of individual stocks}
\label{sec23}

The demand can be quantified as the volume imbalance, \textit{i.e.} the difference between all bought-in volumes and all sold-out volumes in a time interval $t$, 
\begin{equation}
\nu(t)=\sum_{n=1}^{N_{\mathrm{trades}}(t)}v(t;n)\varepsilon(t;n) \ .
\label{eq2.2.2}
\end{equation}
Here, $N_{\mathrm{trades}}(t)$ denotes the number of trades in time interval $t$, and $v(t;n)$ is the unsigned volume for $n$-th trade in $t$. To have, at the same time, many trades in the time intervals $t$ and a long time series $\nu(t)$ in each trading day, we use time intervals $t$ of one minute. It is useful to introduce the local noise intensity~\cite{Plerou2003}, 
\begin{equation}
\Sigma(t)=\left\langle\left|v(t;n)\varepsilon(t;n)-\big\langle v(t;n)\varepsilon(t;n)\big\rangle_n\right|\right\rangle_n \ ,
\label{eq2.2.3}
\end{equation}
which can be understand as the amount of fluctuations around the local average of volume imbalance in a time interval. 

\begin{figure*}[tb]
\begin{center}
\includegraphics[width=1\textwidth]{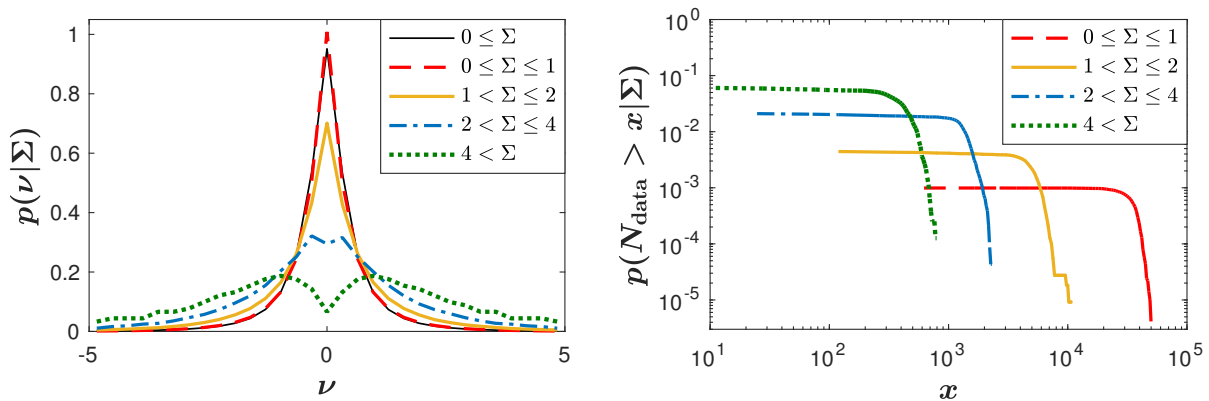}
\caption{Left: the probability density distributions of volume imbalance conditioned on the local noise intensity $p(\nu|\Sigma)$; right: the cumulative probability density distributions of the numbers of data points conditioned on the local noise intensity $p(N_{\mathrm{data}}>x|\Sigma)$ versus variable $x$.}
\vspace*{-0.5cm}
\label{fig21}
\end{center}
\end{figure*}

We investigate the two-phase behavior by examining the distribution of the volume imbalance conditioned on the local noise intensity, $p(\nu|\Sigma)$, as shown in figure~\ref{fig21}. The distributions are found for altogether 496 stocks in S$\&$P 500. To include different stocks on equal footing, we scale out the volatilities by further normalizing $\Sigma(t)$ to zero mean and unit variance. For larger $\Sigma(t)$, the transition from a unimodal distribution to a bimodal distribution appears. Especially, when the $\Sigma(t)>4$, the bimodal distribution is obvious. These large local noise intensity and bimodal distributions are exactly what we are interested in when looking at the copula density of demands between stocks conditioned on the local noise intensity.

\section{Empirical dependencies between stocks}
\label{sec3}

Although there are detailed presentations on copulas in the statistics literature~\cite{Joe1997,Nelsen2007,Wollschlager2015,Chetalova2015b}, we give a short sketch of the concept for the convenience of the reader in section~\ref{sec31}. In section~\ref{sec32}, we illustrate and discuss how the empirical copula densities are estimated. In section~\ref{sec33}, we show the empirical copula density and discuss the asymmetry of tailed dependence of the copula.

\subsection{Copula densities}
\label{sec31}

Let $F_{kl}(x_1,x_2)$ be a joint cumulative distribution of the random variables $x_1$ and $x_2$ with marginal cumulative distributions $F_k(x_1)$ and $F_l(x_2)$, respectively. According to Sklar's theorem\cite{Nelsen2007}, there exists a copula $\mathrm{Cop}_{kl}(q_1,q_2)$ for all quantiles $q_1, q_2 \in[0,1]$ satisfying 
\begin{equation}
F_{kl}(x_1,x_2)=\mathrm{Cop}_{kl}\big(F_k(x_1),F_l(x_2)\big) \ .
\label{eq3.1.1}
\end{equation}
 In terms of the probability density function $f_k(x_1)$ of the random variable $x_1$, the marginal cumulative distribution function $F_k(x_1)$ can be expressed as,
\begin{equation}
F_k(x_1)=\int\limits_{-\infty}^{x_1}f_k(s)ds \ ,
\label{eq3.1.2}
\end{equation}
and analogously for $F_l(x_2)$. The inverse cumulative distribution function $F_k^{-1}(\cdot)$ is known as the quantile function. We thus have
\begin{equation}
q_1=F_k(x_1) \quad \mathrm{and}\quad   x_1=F_k^{-1}(q_1) \ .
\label{eq3.1.3}
\end{equation}
and accordingly for $q_2=F_l(x_2)$. Hence, using equation~(\ref{eq3.1.1}) , the copula can be expressed as the cumulative joint distribution of the quantiles,
\begin{equation}
\mathrm{Cop}_{kl}(q_1,q_2)=F_{kl}\big(F_k^{-1}(q_1),F_l^{-1}(q_2)\big) \ .
\label{eq3.1.4}
\end{equation}
Thus, the dependence structure of random variables is separated from the marginal probability distributions. In other words, the pure dependence structure is measured independently of the particular marginal distribution. The copula density is given as the two-fold derivative
\begin{equation}
\mathrm{cop}_{kl}(q_1,q_2)=\frac{\partial ^2}{\partial q_1\partial q_2}\mathrm{Cop}_{kl}(q_1,q_2) \ 
\label{eq3.1.5} 
\end{equation}
with respect to the quantiles.

\subsection{Empirical estimation of copula densities}
\label{sec32}

To estimate the empirical pairwise copula densities of demands, we first map all observations of volume imbalances $\nu_k(t)$ from stock $k$ to a uniformly distributed time series $q_1(t)$ by
\begin{equation}
q_1(t)=F_k(\nu_k(t))=\frac{1}{T}\sum_{\tau=1}^{T}\Theta\big(\nu_k(t)-\nu_k(\tau)\big)-\frac{1}{2T} \ ,
\label{eq3.2.1}
\end{equation}
where $\Theta (\cdot)$ is the Heaviside step function, and $T$ is the length of the time series. The volume imbalance $\nu_k(t)$ is defined in equation~(\ref{eq2.2.2}). To arrive at generic results, we average over all $L(L-1)/2$ stock pairs,
\begin{equation}
\mathrm{cop}(q_1,q_2)=\frac{2}{L(L-1)}\sum_{k=1}^{L-1}\sum_{l=k+1}^{L}\mathrm{cop}_{kl}(q_1,q_2) \ ,
\label{eq3.2.2}
\end{equation}
where $\mathrm{cop}_{kl}(q_1,q_2)$ is a histogram over two dimensions. The bin size of all these histograms is $\Delta q_1=\Delta q_2=1/20$. Following references~\cite{Wollschlager2015,Chetalova2015b}, we do not use a symmetrized definition of the averaged copula.

One might argue that the empirical copula densities should be averaged over $L(L-1)$ stock pairs by 
\begin{equation}
\mathrm{cop}(q_1,q_2)=\frac{1}{L(L-1)}\sum_{k=1}^{L-1}\sum_{l=k+1}^{L}\Big(\mathrm{cop}_{kl}(q_1,q_2)+\mathrm{cop}_{lk}(q_1,q_2)\Big)\ ,
\label{eq3.2.3}
\end{equation}
which would make the averaged copula densities independent of the order of two stocks in a pair. To clarify the reasons for the choice of definition~(\ref{eq3.2.2}), we first point out that the order of stocks in a pair will not influence the averaged copula densities largely, as we consider the average of copula densities over a large amount of stock pairs. The purpose of averaging is to wash out the individual features of specific stock pairs and to reveal the generic ones. 

\begin{figure*}[tb]
\raggedleft
\begin{minipage}[t]{0.43\textwidth}
\includegraphics[width=1\textwidth]{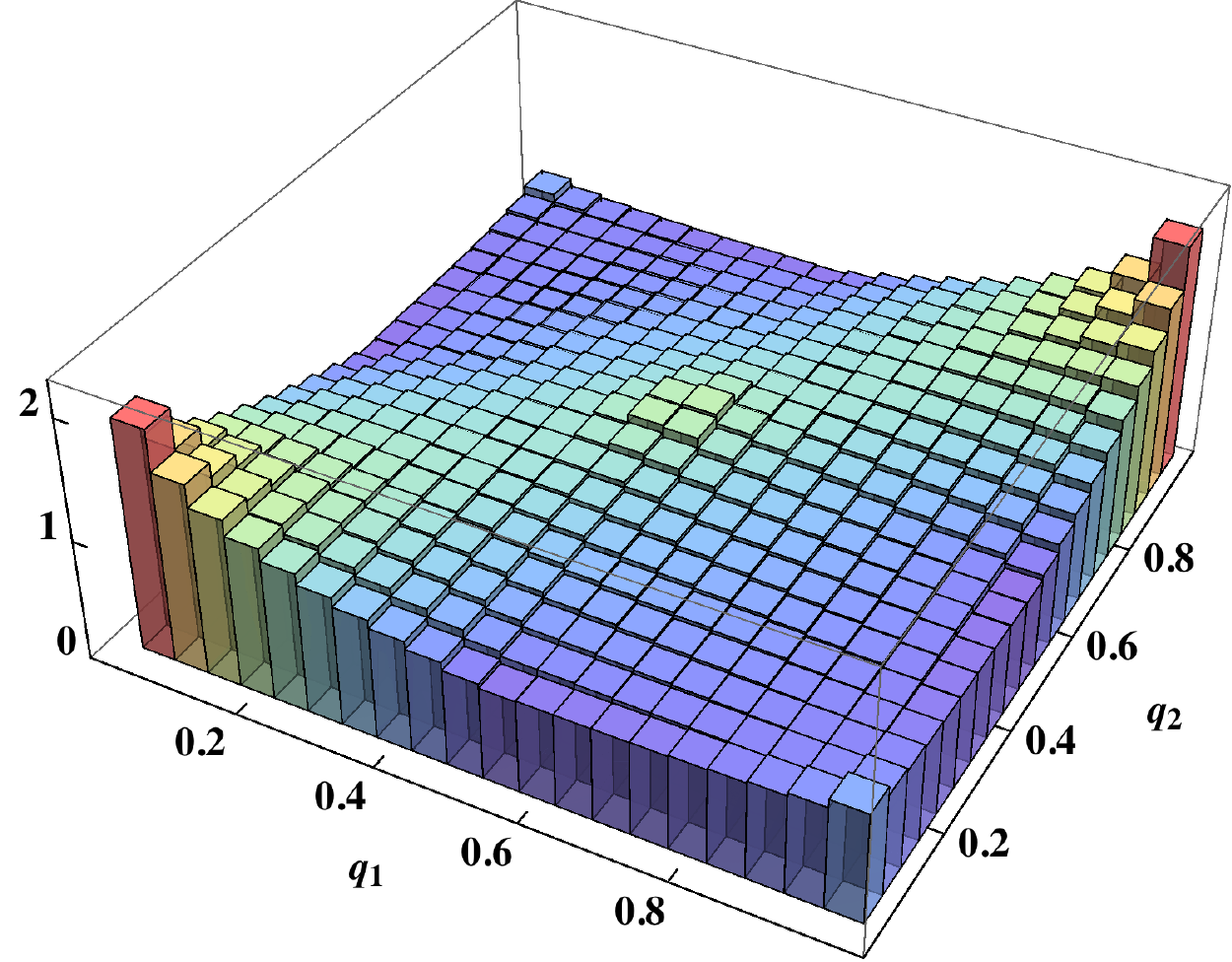}
\end{minipage}
\begin{minipage}[t]{0.43\textwidth}
\includegraphics[width=1\textwidth]{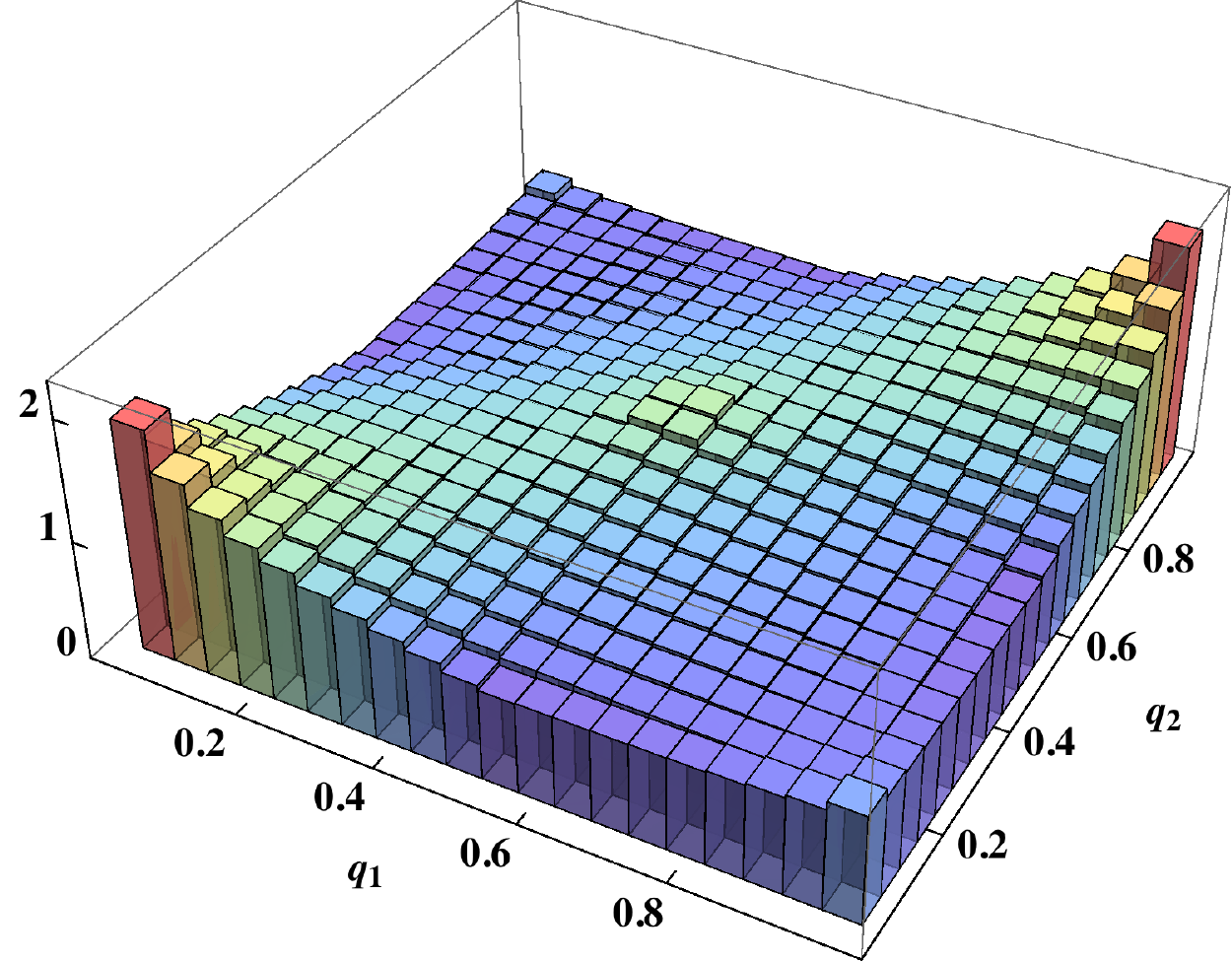}
\end{minipage}
\caption{The empirical copula density $\mathrm{cop}(q_1,q_2)$ of volume imbalances averaged over 4950 stock pairs ($k$, $l$). Left: the order of stocks is preset; right: the order of stocks is shuffled.}
\vspace*{-0.5cm}
\label{fig31}
\end{figure*}

\begin{figure*}[tb]
\raggedleft
\includegraphics[width=0.85\textwidth]{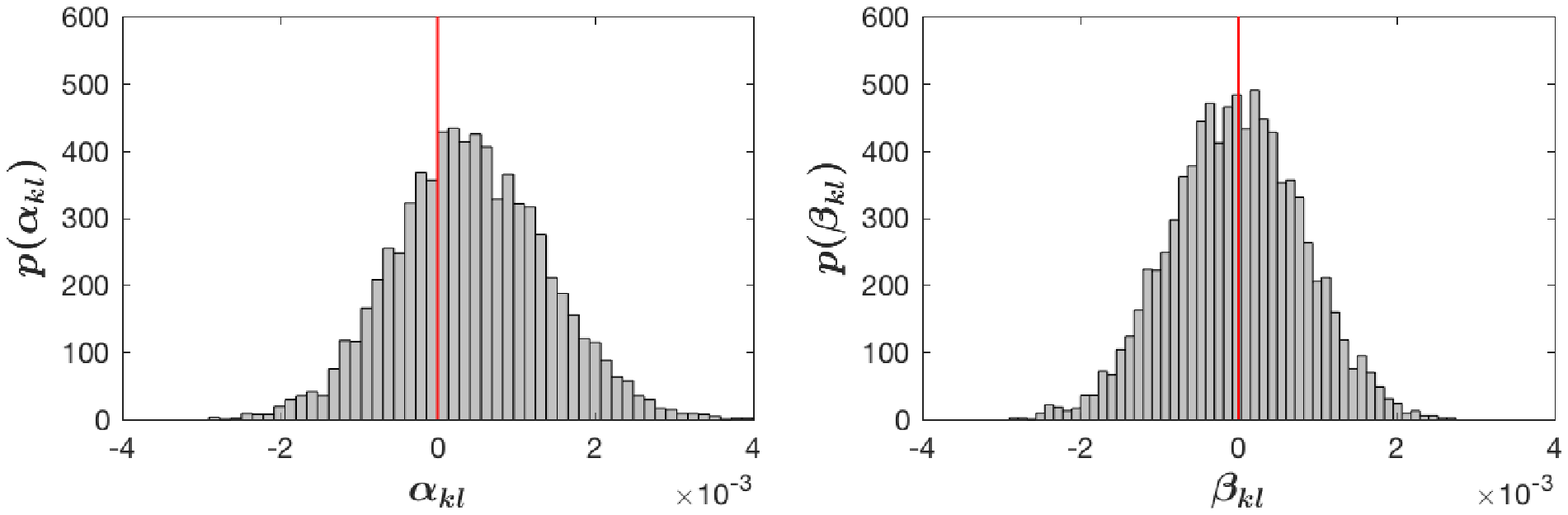}\\
\vspace*{-0.25cm}
\includegraphics[width=0.85\textwidth]{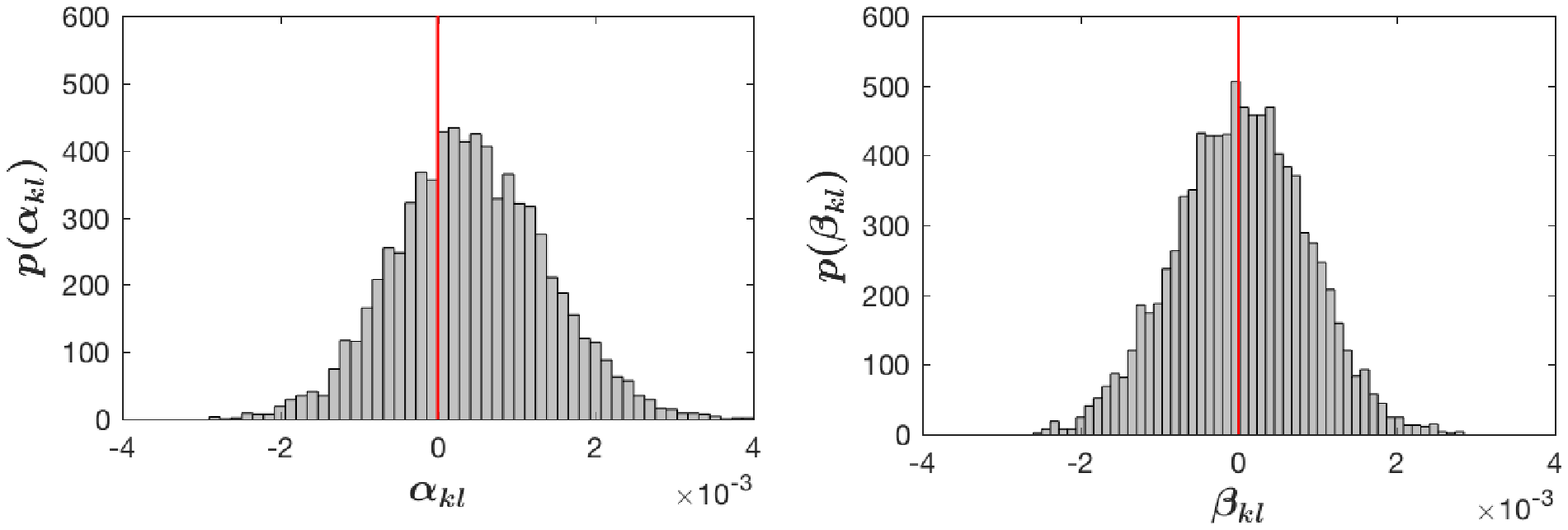}
\vspace*{-0.5cm}
\caption{Histograms of asymmetry values of copula densities for positive dependence $p(\alpha_{kl})$ (left) and for negative dependence $p(\beta_{kl})$ (right) with 4950 stock pairs ($k$, $l$). Top: the order of stocks is preset; bottom: the order of stocks is shuffled. All the histograms are normalized to one.}
\vspace*{-0.5cm}
\label{fig32}
\end{figure*}

When the number of stocks tends to the infinity, \textit{i.e.} $L\rightarrow\infty$, the definitions~(\ref{eq3.2.3}) and (\ref{eq3.2.2}) are equivalent. We calculate the empirical copula density of volume imbalances with the first definition~(\ref{eq3.2.2}), as shown in figure~\ref{fig31}. Here, to facilitate the calculation, we replace $L\rightarrow\infty$ by $L=100$, and total 4950 stock pairs are used. To have a better view of the dependence structure, the tail asymmetries of the copula density are characterized by two quantities, $\alpha_{kl}$ and $\beta_{kl}$, 
\begin{equation}
\alpha_{kl}=\int\limits_{0.8}^{1}dq_1\int\limits_{0.8}^{1}dq_2 ~\mathrm{cop}_{kl}(q_1,q_2)-\int\limits_0^{0.2}dq_1\int\limits_0^{0.2}dq_2 ~\mathrm{cop}_{kl}(q_1,q_2) \ ,
\label{eq3.2.4}
\end{equation}
\begin{equation}
\beta_{kl}=\int\limits_{0}^{0.2}dq_1\int\limits_{0.8}^{1}dq_2 ~\mathrm{cop}_{kl}(q_1,q_2)-\int\limits_{0.8}^{1}dq_1\int\limits_{0}^{0.2}dq_2 ~\mathrm{cop}_{kl}(q_1,q_2) \ ,
\label{eq3.2.5}
\end{equation}
\textit{i.e.}, we look into the corners of size 0.2 times 0.2 in the $(q_1, q_2)$ plane. Thus, $\alpha_{kl}$ describes the asymmetry of positive dependence. A shift away from zero in the histogram of $\alpha_{kl}$ can be seen in figure~\ref{fig32}. However, the asymmetry of the negative dependence, indicated by $\beta_{kl}$, is more significant. An overall symmetric distribution around zero for $\beta_{kl}$ can be found in figure~\ref{fig32}. That implies the averaged $\mathrm{cop}_{kl}(q_1,q_2)$ over 4950 stock pairs ($k$, $l$) is equivalent to the averaged $\mathrm{cop}_{lk}(q_1,q_2)$ over 4950 stock pairs ($l$, $k$). Hence the two definitions~(\ref{eq3.2.2}) and (\ref{eq3.2.3}) are equivalent for all practical purposes. 

The difference of the two definitions~(\ref{eq3.2.2}) and (\ref{eq3.2.3}) lies in whether or not the order of stocks influences the dependence structure of average copula density. We therefore shuffle the order of stocks and recalculate the copula density with definition~(\ref{eq3.2.2}). The results in figures~\ref{fig31} and \ref{fig32} do not change too much compared to the original ones with preset order of stocks as listed in \ref{appA}.

We thus employ the definition~(\ref{eq3.2.2}) to average the empirical copula densities rather than the definition~(\ref{eq3.2.3}).

\subsection{Empirical copula densities}
\label{sec33}

From the empirical copula densities of volume imbalances in figure~\ref{fig31}, strong dependencies of large demands between stocks can be inferred, positive as well as negative ones. The positive demands mean that the buyer-initiated trades dominate in the market. The negative demands correspond to supplies of volumes, \textit{i.e.}, seller-initiated trades dominate. Thus, either the large supplies or the large demands between stocks exhibit strong, positive dependencies. In contrast, the dependencies between large supply of one stock and large demand of another stock, \textit{i.e.} the negative dependencies, also exist, but are not as pronounced.

As we have seen, the negative dependencies are almost symmetric for the average copula densities, but the positive dependencies are not. Once more, the asymmetry of the $\alpha_{kl}$ distribution in figure~\ref{fig31} is important, as it implies a stronger dependence of demands than of supplies. To further quantify the asymmetry of distributions, we introduce the skewness, defined as
\begin{equation}
\mathrm{skewness}=\frac{\langle (x-\mu)^3\rangle}{\sigma^3} \ ,
\label{eq3.3.1}
\end{equation}
where $\mu$ is the mean of $x$, and $\sigma$ is the standard deviation of $x$. Here, $x$ stands for $\alpha_{kl}$ and $\beta_{kl}$, respectively. We thus measure the skewness of the distributions, listed in table~\ref{tab1}, where the one for $\alpha_{kl}$ is 0.0977. This suggests that from a large trade of one stock, it is more likely to find similar trades of other stocks, where the possibility of buy trades is higher than the possibility of sell trades. When the traded volumes are much larger than the market depth, these buy trades will push the prices up~\cite{Plerou2005,Wang2016a,Wang2016b}. In financial markets, the persistent raising of prices of most stocks indicates a bull market. Consequently, the asymmetry of positive dependencies suggests the traders are more optimistic expecting a bull market.

\begin{table}[b]
\caption{\label{tab1}The skewness of distribution of asymmetries}
\begin{center}
\begin{footnotesize}
\begin{tabular}{@{\hskip 0.2in}c@{\hskip 0.2in}c@{\hskip 0.2in}c@{\hskip 0.2in}c@{\hskip 0.45in}c@{\hskip 0.45in}c@{\hskip 0.45in}c@{\hskip 0.2in}}
\hlineB{2}
	& \makecell{$p(\lambda_{kl})$ preset} & \makecell{$p(\lambda_{kl})$ shuffled} & $p(\lambda_{kl}^{(\mathrm{ss})})$ & $p(\lambda_{kl}^{(\mathrm{ll})})$ & $p(\lambda_{kl}^{(\mathrm{sl})})$ & $p(\lambda_{kl}^{(\mathrm{ls})})$ \\
\hline
\makecell{$\lambda=\alpha$} 	& 0.0977	&0.0977	&0.0665	&0.1247	&-0.0351	&0.0008	\\
\makecell{$\lambda=\beta$} & -0.0319	& -0.0257	 &-0.0373	&-0.0473	&-0.0186	&-0.0140	\\ 		
\hlineB{2}
\end{tabular}
\end{footnotesize}
\end{center}
\end{table}

\section{Comparison of two models with the empirical copula density}
\label{sec4}

To explain the empirical dependence between stocks, we fit the empirical copula density with two functions, a bivariate $\mathcal{K}$ copula density function and a Gaussian copula density function. Since the two copula density functions are discussed in references~\cite{Chetalova2015b,Wollschlager2015}, we only shortly introduce them in sections~\ref{sec41} and \ref{sec42}, respectively. We then compare them with the empirical results in section~\ref{sec43}.

\subsection{Bivariate $\mathcal{K}$ copula density}
\label{sec41}

A $K$ component vector $\mathbf{r}=\big(r_1,... ,r_K\big)$ with elements $r_k$, $k=1,... ,K$, normalized to zero mean and unit variance, follows a multivariate $\mathcal{K}$ distribution~\cite{Schmitt2013,Chetalova2015c}, if its probability density is given by
\begin{eqnarray} \nonumber
\langle g \rangle(\mathbf{r} | C,N)&=&\frac{1}{2^{N/2+1}\Gamma(N/2)\sqrt{\mathrm{det}(2\pi C/N)}}\frac{\mathcal{K}_{(K-N)/2}\left(\sqrt{N\mathbf{r}^{\dagger}C^{-1}\mathbf{r}}\right)}{\sqrt{N\mathbf{r}^{\dagger}C^{-1}\mathbf{r}}^{(K-N)/2}} \\
&=&\frac{1}{(2\pi)^K\Gamma(N/2)\sqrt{\mathrm{det}C}}\int\limits_0^\infty dzz^{\frac{N}{2}-1}e^{-z}\sqrt{\frac{\pi N}{z}}^K \mathrm{exp}\left(-\frac{N}{4z}\mathbf{r}^\dagger C^{-1}\mathbf{r}\right) \ .
\label{eq4.1.1}
\end{eqnarray}
The notation $\langle g \rangle$ indicates that this distribution results from a random matrix average to model non-stationary, \textit{i.e.}, fluctuating covariance or correlation matrices with a mean value $C$. The parameter $N$ measures the strength of these fluctuations, $1/N$ can be viewed as the corresponding variance. $\mathcal{K}_{m}$ is the modified Bessel function of the second kind of order $m$. In the present content $\mathbf{ r}$ is a vector of returns, which  are time series $r_k=r_k(t), t=1,\cdots,T$. The distribution~(\ref{eq4.1.1}) is assumed to hold for each time $t$. It is worth mentioning that the parameter $N$ is different from $N_{\mathrm{trades}}(t)$ in equation~(\ref{eq2.2.2}), which represents the number of trades in the time interval $t$. In the bivariate case $K=2$, the joint pdf of the $\mathcal{K}$ distribution reads,
\begin{eqnarray}
f(x_1,x_2)&=&\frac{1}{\Gamma(N/2)}\int\limits_0^\infty dzz^{\frac{N}{2}-1}e^{-z}\frac{N}{4\pi z}\frac{1}{\sqrt{1-c^2}}\mathrm{exp}\left(-\frac{N}{4z}\frac{x_1^2-2cx_1x_2+x_2^2}{1-c^2}\right) \ , 
\label{eq4.1.2}
\end{eqnarray}
with the correlation matrix
\begin{equation}
C=\left[\begin{array}{cc}1 & c \\c & 1\end{array}\right] \ ,
\label{eq4.1.3}
\end{equation}
which only depends on one correlation coefficient $c$. By integrating $f(x_1,x_2)$ over the whole range of $x_2$, we can obtain the marginal distribution density,   
\begin{eqnarray}\nonumber
f_k(x_1)&=&\int\limits_{-\infty}^{\infty}dx_2 f(x_1,x_2) \\
&=&\frac{1}{\Gamma(N/2)}\int\limits_0^{\infty}dz z^{\frac{N}{2}-1}e^{-z}\sqrt{\frac{N}{4\pi z}}exp\left(-\frac{N}{4z}x_1^2\right) \ ,
\label{eq4.1.4}
\end{eqnarray}
and analogously for $f_l(x_2)$. Further, the integral of the probability density function yields the marginal cumulative distribution,
\begin{eqnarray} \nonumber
F_k(x_1)&=&\int\limits_{-\infty}^{x_1} d\xi f_k(\xi) \\
&=&\frac{1}{\Gamma(N/2)}\int\limits_0^{\infty}dz z^{\frac{N}{2}-1}e^{-z}\int\limits_{-\infty}^{x_1} d\xi\sqrt{\frac{N}{4\pi z}}exp\left(-\frac{N}{4z}\xi^2\right) \ ,
\label{eq4.1.5}
\end{eqnarray}
and $F_l(x_2)$ accordingly.
With equations~(\ref{eq3.1.4}) and (\ref{eq3.1.5}), the copula density function can be derived as
\begin{equation} 
\mathrm{cop}_{c,N}^\mathcal{K}(q_1,q_2)=\frac{f\big(F_k^{-1}(q_1),F_l^{-1}(q_2)\big)}{f_k\big(F_k^{-1}(q_1)\big)f_l\big(F_l^{-1}(q_2)\big)} \ .
\label{eq4.1.6}
\end{equation}
A more detailed discussion of the bivariate $\mathcal{K}$ copula is given in reference~\cite{Chetalova2015b}.

\subsection{Gaussian copula density}
\label{sec42}

Here, one assumes that the random variables $x_1$ and $x_2$, normalized to zero mean and unit variance, follow a bivariate normal distribution with a correlation coefficient $c$. The bivariate cumulative normal distribution of $x_1$ and $x_2$ is given by
\begin{equation}
F(x_1,x_2)=\int\limits_{-\infty}^{x_1}\int\limits_{-\infty}^{x_2}\frac{1}{2\pi \sqrt{1-c^2}}\mathrm{exp}\left(-\frac{y_1^2+y_2^2-2cy_1y_2}{2(1-c^2)} \right) dy_2dy_1 \ .
\label{eq4.2.1}
\end{equation}
Hence, the marginal cumulative normal distribution of $x_1$ is 
\begin{equation}
F_k(x_1)=\int\limits_{-\infty}^{x_1}\frac{1}{\sqrt{2\pi}}\mathrm{exp}\left(-\frac{y_1^2}{2} \right) dy_1 \ ,
\label{eq4.2.2}
\end{equation} 
and analogously for $F_l(x_2)$. Using  equations~(\ref{eq4.2.1}), (\ref{eq4.2.2}) and (\ref{eq3.1.4}), we find an explicit expression of the Gaussian copula density 
\begin{eqnarray} \nonumber
\mathrm{cop}_{c}^G(q_1,q_2)&=&\frac{\partial ^2}{\partial q_1\partial q_2}F\big(F_k^{-1}(q_1), F_l^{-1}(q_2)\big) \\
&=&\frac{1}{\sqrt{1-c^2}}\mathrm{exp}\left(-\frac{c^2F_k^{-1}(q_1)^2+c^2F_l^{-1}(q_2)^2-2cF_k^{-1}(q_1)F_l^{-1}(q_2)}{2(1-c^2)}\right) \ 
\label{eq4.2.3}
\end{eqnarray}
by carrying out the partial derivatives in equation~(\ref{eq3.1.5}).

\subsection{Fits}
\label{sec43}

\begin{figure*}[tb]
\raggedleft
\includegraphics[width=0.8\textwidth]{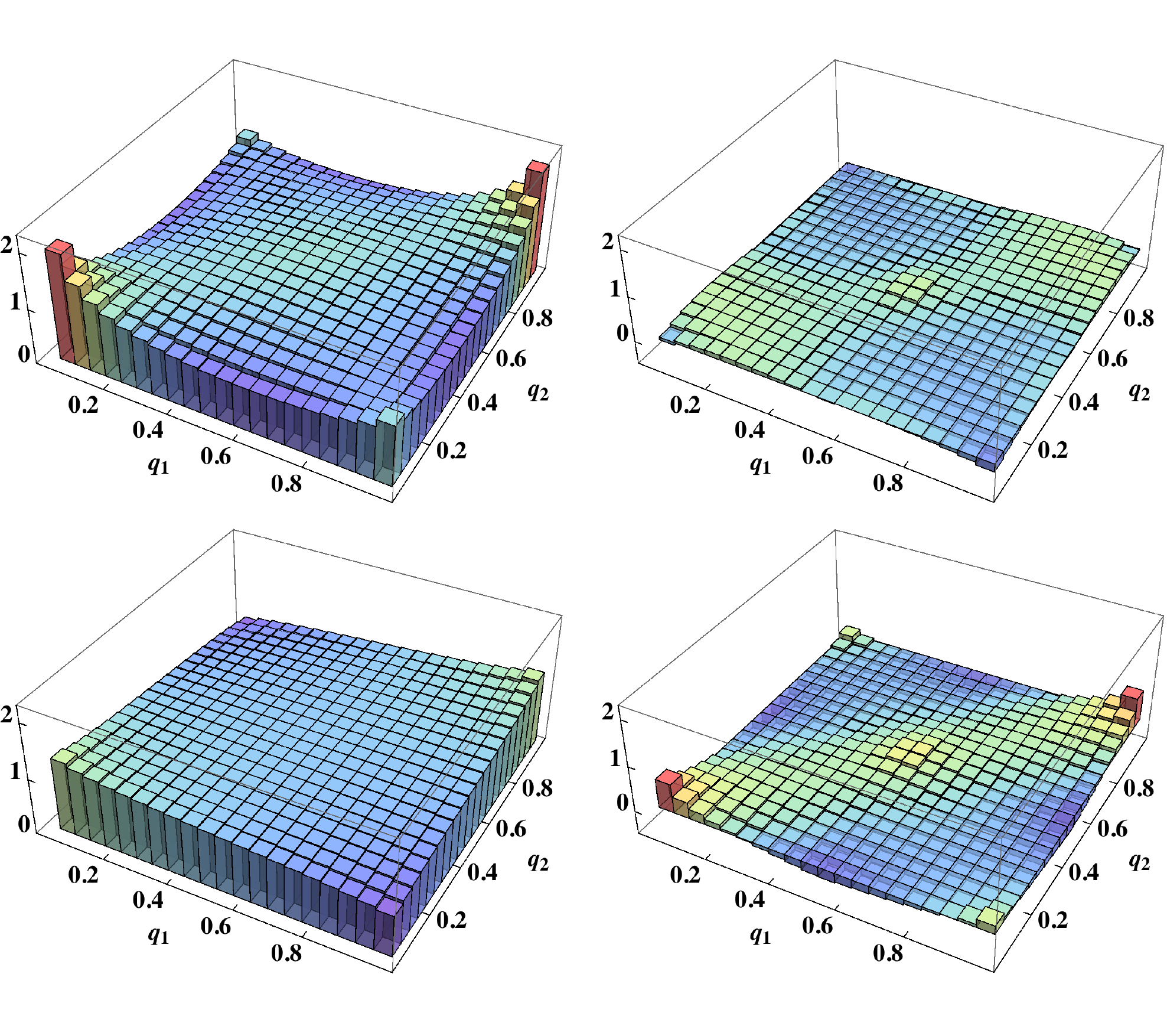}
\vspace*{-0.7cm}
\caption{$\mathcal{K}$ copula density $\mathrm{cop}_{\bar{c},N}^{\mathcal{K}}(q_1,q_2)$ with $\bar{c}=0.10$ and $N=6.72$ (left, top). The error between the empirical copula density and the $\mathcal{K}$ copula density, defined as $\mathrm{cop}(q_1, q_2)-\mathrm{cop}_{\bar{c},N}^{\mathcal{K}}(q_1, q_2)$, (right, top). Gaussian copula density $\mathrm{cop}_{\bar{c}}^{G}(q_1,q_2)$ with $\bar{c}=0.10$ (left, bottom). The error between the empirical copula density and the Gaussian copula density, defined as $\mathrm{cop}(q_1, q_2)-\mathrm{cop}_{\bar{c}}^{G}(q_1, q_2)$, (right, bottom). }
\vspace*{-0.3cm}
\label{fig41}
\end{figure*}

To fit the empirical copula, we first work out the average correlation coefficient $\bar{c}=0.10$ by averaging over $L(L-1)/2$ stock pairs for the $L=100$ corresponding to 100 stocks listed in \ref{appA}. In the $\mathcal{K}$ copula density function in equation~(\ref{eq4.1.6}), the correlation coefficient $c$ is replaced by $\bar{c}$. Thus, in equation~(\ref{eq4.1.6}), only the free parameter $N$ needs to be fitted. By minimizing the squared difference between the empirical copula density and the model copula density, we find $N=6.72$. With the same $\bar{c}$, we also carry out this comparison using the Gaussian copula density function in equation~(\ref{eq4.2.3}). To quantify the goodness of fit, we work out the difference between the empirical copula density and the model copula density. Figure~\ref{fig41} shows the two fits and the difference between data and model. The Gaussian copula density differs from the empirical one by a large extent. In particular, the tailed dependencies are poorly captured. In contrast, the $\mathcal{K}$ copula density exhibits a good fit to the empirical result, as it works much better for the tails. This supports previous studies in which the $\mathcal{K}$ distribution was found to give good descriptions of multivariate data subject to nonstationarities~\cite{Schmitt2013,Schmitt2014,Chetalova2015c,Chetalova2015a}.

\section{Influence of local fluctuations on dependencies}
\label{sec5}

In section~\ref{sec51}, We discuss a method to analyze the conditional copula density. In section~\ref{sec52}, we define copula densities conditioned on the local noise intensity and discuss the influence of large local fluctuations on the dependence of volume imbalances between stocks. In section~\ref{sec53}, we give an explanation of this influence with respect to the cross-correlation of volume imbalances. In section~\ref{sec54}, we investigate the influence of large local fluctuations on the asymmetries of tailed dependencies. 

\subsection{Feasibility of our method}
\label{sec51}

We work out the cumulative probability densities of the numbers of data points $N_{\mathrm{data}}$ for the four ranges of noises in figure~\ref{fig21}. A data point gives a volume imbalance as well as a corresponding local noise intensity in the time interval of one minute. As seen in figure~\ref{fig21}, for larger numbers of data points, it is less possible to observe the bimodal distribution. In particular, for some stocks, the bimodal distribution with $\Sigma>4$ results from only several dozens of data points. When considering the conditional dependencies of demands between two individual stocks that have bimodal marginal distributions, however, these data points are not sufficient to have access to the better statistical property. We thus employ the following method to measure the influence of large local noise intensity. First, we work out the conditional copula density, excluding 50 data points with the largest local noise intensity from both stocks or either stock of a pair. We find little difference between the copula densities excluding 10, 50 and 100 such data points, respectively. However, due to data points that result in a unimodal distribution, enlarging the number of such data points to more than 100 will make the copula density different. Next, we subtract that conditional copula density from the corresponding unconditional one including all data points. The difference between them is the part induced by the large local noise intensity. 

\subsection{Influence on the dependence structure}
\label{sec52}

\begin{figure*}[p]
\raggedleft
\vspace*{-0.3cm}
\includegraphics[width=0.77\textwidth]{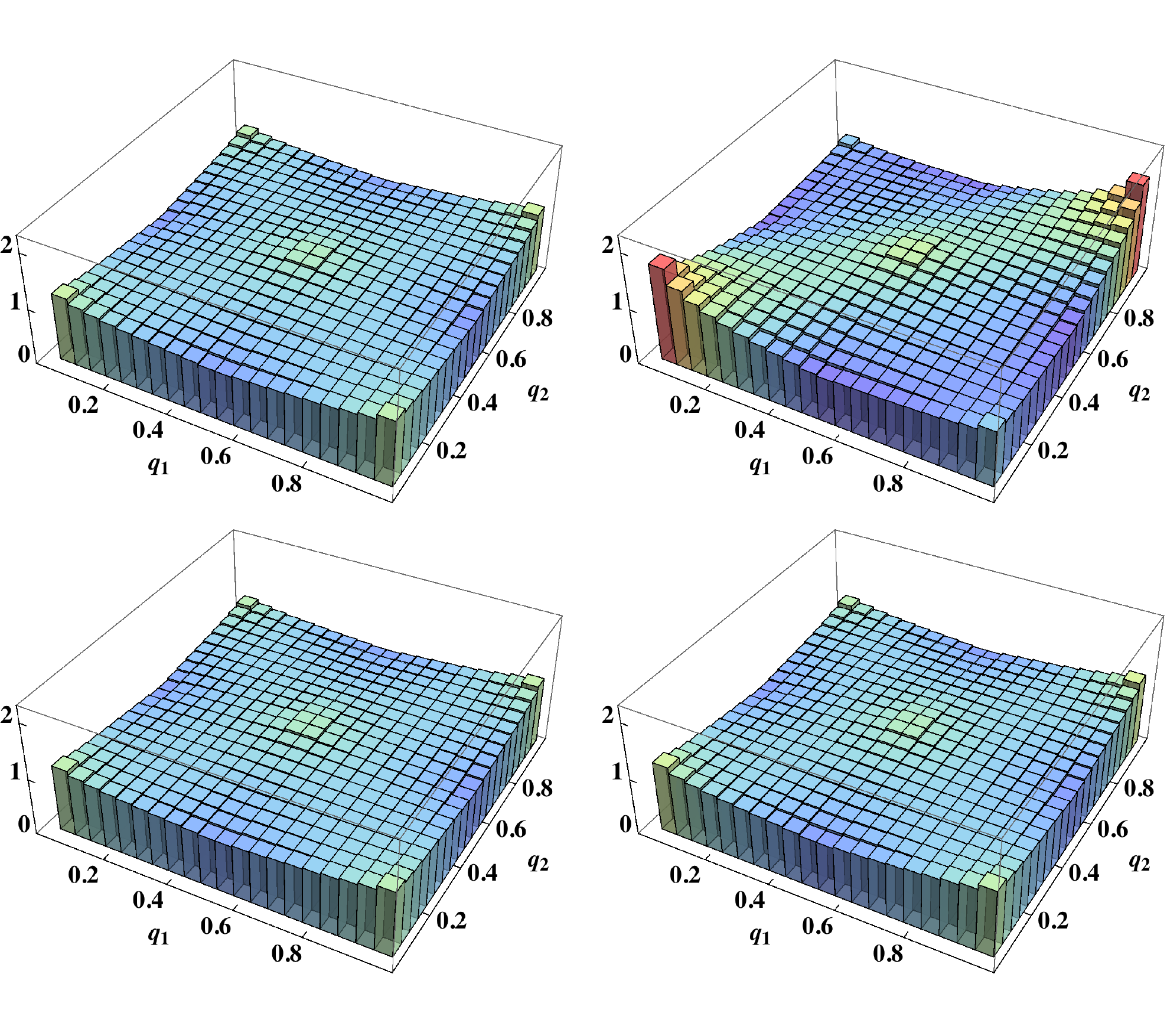}
\vspace*{-0.7cm}
\caption{Empirical copula densities conditioned on the local noise intensity $\mathrm{cop}^{(\mathrm{ss})}(q_1,q_2)$ (left, top), $\mathrm{cop}^{(\mathrm{ll})}(q_1,q_2)$ (right, top), $\mathrm{cop}^{(\mathrm{sl})}(q_1,q_2)$ (left, bottom), and $\mathrm{cop}^{(\mathrm{ls})}(q_1,q_2)$  (right, bottom).}
\vspace*{-0.5cm}
\label{fig51}
\end{figure*}

We now condition the empirical copula densities on the local noise intensity $\Sigma$. The conditional copula densities are worked out by excluding the first 50 data points with the largest or smallest local noise intensity. The exclusion of data points with extremely small local noise intensity is to rule out the construed effect that the large change of dependence structure is randomly induced by excluding any kind of data points. Let $\Sigma_{k,\mathrm{max}}$ denote the minimum of the first 50 data points with the extremely large local noise intensity for stock $k$, and $\Sigma_{k,\mathrm{min}}$ the maximum of the first 50 data points with extremely small local noise intensity for this stock. We write the conditional copula densities as
\begin{eqnarray}\nonumber
\mathrm{cop}^{(\mathrm{ss})}(q_1,q_2)&=&\mathrm{cop}\left(q_1,q_2\big|\Sigma_k<\Sigma_{k,\mathrm{max}},\Sigma_l<\Sigma_{l,\mathrm{max}}\right) \ ,\\ \nonumber
\mathrm{cop}^{(\mathrm{ll})}(q_1,q_2)&=&\mathrm{cop}\left(q_1,q_2\big|\Sigma_k>\Sigma_{k,\mathrm{min}},\Sigma_l>\Sigma_{l,\mathrm{min}}\right) \ , \\ \nonumber
\mathrm{cop}^{(\mathrm{sl})}(q_1,q_2)&=&\mathrm{cop}\left(q_1,q_2\big|\Sigma_k<\Sigma_{k,\mathrm{max}},\Sigma_l>\Sigma_{l,\mathrm{min}}\right) \ , \\
\mathrm{cop}^{(\mathrm{ls})}(q_1,q_2)&=&\mathrm{cop}\left(q_1,q_2\big|\Sigma_k>\Sigma_{k,\mathrm{min}},\Sigma_l<\Sigma_{l,\mathrm{max}}\right) \ .
\label{eq5.1.1}
\end{eqnarray}
Here, $\Sigma_k$ and $\Sigma_l$ are the local noise intensity for stock $k$ and stock $l$, respectively. Furthermore, $\mathrm{cop}^{(\mathrm{ss})}(q_1,q_2)$ indicates that the copula density results from the quantiles $q_1$ and $q_2$ with small local noise intensity, while $\mathrm{cop}^{(\mathrm{ll})}(q_1,q_2)$ represents the opposite case. Similarly, $\mathrm{cop}^{(\mathrm{sl})}(q_1,q_2)$ denotes the copula density from the quantiles $q_1$ with small local noise intensity and the quantiles $q_2$ with the large local noise intensity, and vice versa for $\mathrm{cop}^{(\mathrm{ls})}(q_1,q_2)$. We show the four types of conditional copula densities in figure~\ref{fig51}. Only $\mathrm{cop}^{(\mathrm{ll})}(q_1,q_2)$ reveals strongly positive dependencies, the dependencies in the other copula densities are nearly uniform at the corners and the centres.

\begin{figure*}[htbp]
\raggedleft
\includegraphics[width=0.77\textwidth]{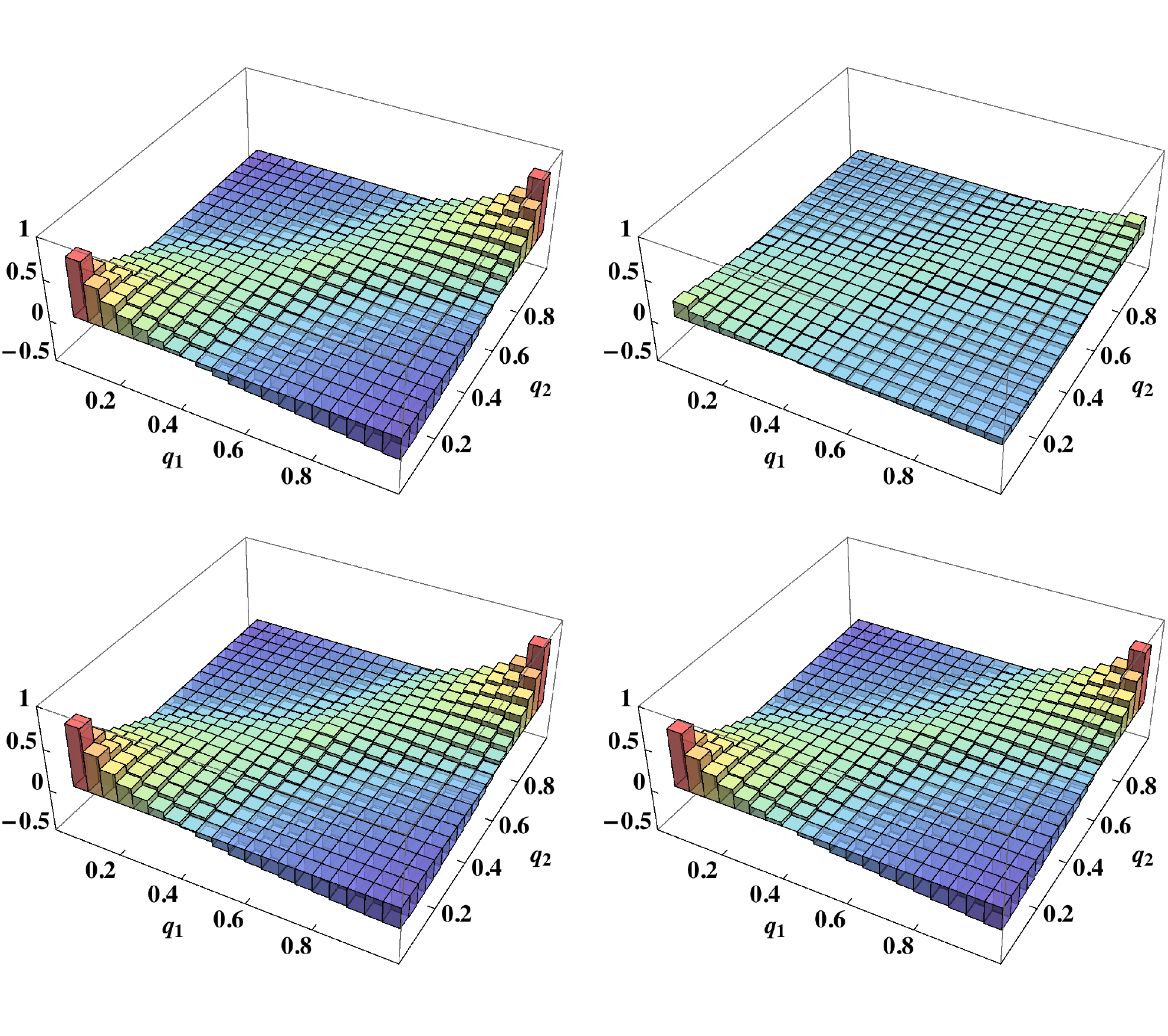}
\vspace*{-0.7cm}
\caption{The influences of the local noise intensity on the copula density $\Delta \mathrm{cop}^{(\mathrm{ll})}(q_1,q_2)$ (left, top), $\Delta \mathrm{cop}^{(\mathrm{ss})}(q_1,q_2)$ (right, top), $\Delta \mathrm{cop}^{(\mathrm{ls})}(q_1,q_2)$ (left, bottom), and $\Delta \mathrm{cop}^{(\mathrm{sl})}(q_1,q_2)$  (right, bottom).}
\vspace*{-0.5cm}
\label{fig52}
\end{figure*}

To study the influence of large local fluctuations, indicated by the local noise intensity, we look at the difference between the unconditional and the conditional copula densities,
\begin{eqnarray} \nonumber
\Delta \mathrm{cop}^{(\mathrm{ll})}(q_1,q_2)&=&\mathrm{cop}(q_1,q_2)-\mathrm{cop}^{(\mathrm{ss})}(q_1,q_2) \ ,\\ \nonumber
\Delta \mathrm{cop}^{(\mathrm{ss})}(q_1,q_2)&=&\mathrm{cop}(q_1,q_2)-\mathrm{cop}^{(\mathrm{ll})}(q_1,q_2) \ , \\ \nonumber
\Delta \mathrm{cop}^{(\mathrm{ls})}(q_1,q_2)&=&\mathrm{cop}(q_1,q_2)-\mathrm{cop}^{(\mathrm{sl})}(q_1,q_2) \ ,\\
\Delta \mathrm{cop}^{(\mathrm{sl})}(q_1,q_2)&=&\mathrm{cop}(q_1,q_2)-\mathrm{cop}^{(\mathrm{ls})}(q_1,q_2) \ .
\label{eq5.1.2}
\end{eqnarray}
The unconditional copula density $\mathrm{cop}(q_1,q_2)$ is worked out with all data points. As shown in figure~\ref{fig52}, the extremely small local fluctuations from two stocks have a very slight effect on the positive dependencies of the copula density. This effect is quantified by $\Delta \mathrm{cop}^{(\mathrm{ss})}(q_1,q_2)$. However, the extremely large local fluctuations present in either stock or both stocks of a pair not only enhance the positive dependencies, but also suppress the negative dependencies of the copula densities. The degrees of enhancing and suppressing are measured by $\Delta \mathrm{cop}^{(\mathrm{sl})}(q_1,q_2)$, $\Delta \mathrm{cop}^{(\mathrm{ls})}(q_1,q_2)$, and $\Delta \mathrm{cop}^{(\mathrm{ll})}(q_1,q_2)$, respectively. Comparing the effects of the extremely large and small local fluctuations, we find that the construed effect is absent in the change of dependencies due to large local fluctuations. In the copula densities, the lower corner along the positive diagonal, corresponding to the negative volume imbalances, reveals the dependence of supplies between stocks, while the upper corner along the positive diagonal, corresponding to the positive volume imbalances, reveals the dependence of demands. Combining figures~\ref{fig51} and \ref{fig52}, we find that the extremely large local fluctuations in either stock of a pair are crucial to prompt the strong dependence between demands or supplies. A possible interpretation might be as following: An extremely large trade may either be random or include useful information. In any case, the extremely large trade pushes the price up for a large demand or drops the price down for a large supply. Increase of the price may induce higher expectation for the raising of the price or induce herding behavior~\cite{Scharfstein1990,Bikhchandani2000} of trading, leading more volumes to be bought. Analogously, drop of the price leads to more volumes being sold. Due to the correlations between stocks~\cite{Kullmann2002,Plerou2002}, the effect of a large demand or supply of one stock is very likely to spread to another stock and induce the similar behavior for the volumes to be bought or sold. The presence of such large trades in both stocks of a pair causes, on the one hand, large local fluctuations in the stocks, and, on the other hand, mutual dependence of demands or supplies in the considered stock pair.

\subsection{Correlations induced by large local fluctuations}
\label{sec53}

\begin{figure*}[tbp]
\begin{center}
\includegraphics[width=1\textwidth]{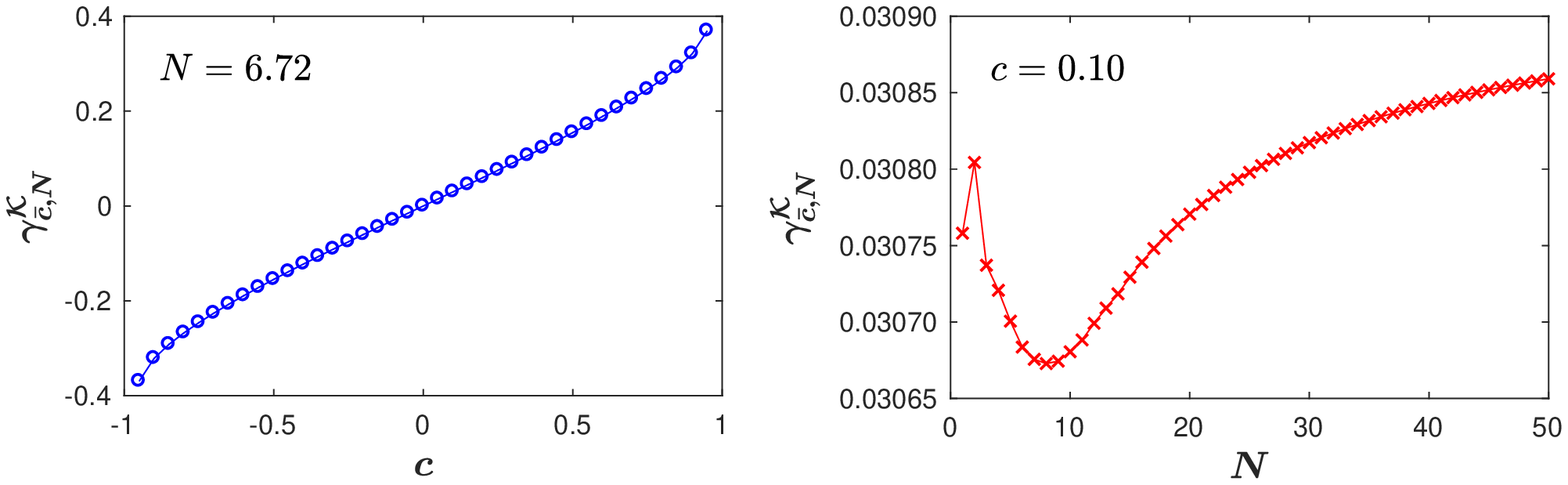}
\vspace*{-0.9cm}
\caption{Left: the dependence of $\gamma_{\bar{c},N}^{\mathcal{K}}$ on the correlation coefficient $c$, where $N=6.72$; right: the dependence of $\gamma_{\bar{c},N}^{\mathcal{K}}$ on the parameter $N$, where $c=0.10$. Here, $\gamma_{\bar{c},N}^{\mathcal{K}}$ is the difference between positive and negative dependencies for the bivariate $\mathcal{K}$ copula density. The ranges of vertical axes for two subgraphs are different.}
\vspace*{-0.5cm}
\label{fig53}
\end{center}
\end{figure*}

As shown in the top row of figure~\ref{fig51}, the change due to large local fluctuations are mainly visible in the positive and negative corners of the copula density. To quantify how such fluctuations affect the dependence structure, we define the difference between positive and negative dependencies of demands for a stock pair ($k$, $l$) as
\begin{eqnarray} \nonumber
\gamma_{kl}&=&\Big(\int\limits_{0.8}^{1}dq_1\int\limits_{0.8}^{1}dq_2 ~\mathrm{cop}_{kl}(q_1,q_2)+\int\limits_0^{0.2}dq_1\int\limits_0^{0.2}dq_2 ~\mathrm{cop}_{kl}(q_1,q_2)\Big) \\
&-&\Big(\int\limits_{0}^{0.2}dq_1\int\limits_{0.8}^{1}dq_2 ~\mathrm{cop}_{kl}(q_1,q_2)+\int\limits_{0.8}^{1}dq_1\int\limits_{0}^{0.2}dq_2 ~\mathrm{cop}_{kl}(q_1,q_2) \Big)\ ,
\label{eq5.2.1}
\end{eqnarray} 
where the terms in the brackets do not coincide with $\alpha_{kl}$ and $\beta_{kl}$ in equations~(\ref{eq3.2.4}) and (\ref{eq3.2.5}). However, the amount of data points is not large enough to empirically analyze $\gamma_{kl}$. Rather, we resort to the $\mathcal{K}$ copula density~(\ref{eq4.1.6}) which, as we have shown, describes the data well. Hence, we replace $\mathrm{cop}_{kl}(q_1,q_2)$ by $\mathrm{cop}_{\bar{c},N}^{\mathcal{K}}(q_1,q_2)$ in definition~(\ref{eq4.1.6}) with $\gamma_{\bar{c},N}^{\mathcal{K}}$ instead of $\gamma_{kl}$. Using equation~(\ref{eq4.1.6}), we calculate $\gamma_{\bar{c},N}^{\mathcal{K}}$ as a function of $c$ for a given $N$ and vice versa, respectively, as shown in figure~\ref{fig53}. The two given values $N=6.72$ and $c=0.10$ are from the fit to the empirical copula. We find that the difference between positive and negative dependencies of demands is drastically affected by the correlation coefficient $c$ rather than by the parameter $N$. This leads us to dissect the cross-correlation of the volume imbalance between stocks $k$ and $l$, 
\begin{eqnarray} \nonumber
\mathrm{corr}\big(\nu_k(t),\nu_l(t)\big)&=&\big\langle\nu_k(t)\nu_l(t)\big\rangle_t \\  \nonumber
&=& \big\langle p_k^+(t) |\nu_k(t)|p_l^+(t)|\nu_l(t)| +p_k^-(t)|\nu_k(t)|p_l^-(t)|\nu_l(t)|  \\ \nonumber
&&	-p_k^+(t) |\nu_k(t)|p_l^-(t)|\nu_l(t)| -p_k^-(t)|\nu_k(t)|p_l^+(t)|\nu_l(t)|  \big\rangle_t \\ 
&=&\big\langle P_{kl}(t)|\nu_k(t)||\nu_l(t)|\big\rangle_t  \ .
\label{eq5.2.2}
\end{eqnarray}
Here, $p_k^+(t)$ is the probability that a surplus of volumes is bought for stock $k$ in the time interval $t$, corresponding to a positive volume imbalance of stock $k$, and $p_k^-(t)$ is the probability that a surplus of volumes is sold, corresponding to a negative volume imbalance. Importantly, we have $p_k^+(t)+p_k^-(t)=1$. The quantity $P_{kl}(t)$ introduced in equation~(\ref{eq5.2.2}) can be written as
\begin{eqnarray} \nonumber
P_{kl}(t)&=&p_k^+(t)p_l^+(t) +p_k^-(t)p_l^-(t)-p_k^+(t)p_l^-(t)-p_k^-(t)p_l^+(t) \\
&=& 4p_k^+(t)p_l^+(t)-2p_k^+(t)-2p_l^+(t)+1 \ ,
\label{eq5.2.3}
\end{eqnarray}
and may be interpreted as effective weight referring to the volume imbalances of both stocks at each time step $t$. The value of $P_{kl}(t)$ is bound between -1 and 1.

In  reference~\cite{Potters2003}, Potters and Bouchaud (BP) have demonstrated that the local noise intensity
\begin{equation}
\widetilde{\Sigma}(t)=\Big\langle\big(v(t;n)\varepsilon(t;n)-\langle v(t;n)\varepsilon(t;n)\rangle_n\big)^2\Big\rangle_n \ ,
\label{eq5.2.4}
\end{equation}
and the square of volume imbalances are positively correlated, 
\begin{eqnarray} \nonumber
\left\langle \widetilde{\Sigma}(t) \nu^2(t)\right\rangle&=&(N_\mathrm{trades}-1)\Big(\left\langle v^4(t;n)\right\rangle-3\left\langle v^2(t;n)\right\rangle^2\Big) \\ 
&&+(1-\frac{3}{N_\mathrm{trades}})\sum_{n_i\neq n_j=1}^{N_\mathrm{trades}}\left\langle v^2(t;n_i)v^2(t;n_j)\right\rangle,
\end{eqnarray}
if the traded volumes have fat tails, \textit{i.e.}, $\left\langle v^4(t;n)\right\rangle>3\left\langle v^2(t;n)\right\rangle^2$, and/or are positively correlated, \textit{i.e.}, $\left\langle v^2(t;n_i)v^2(t;n_j)\right\rangle\geq 0$. They neglect the fluctuation of the number of trades $N_\mathrm{trades}=N_\mathrm{trades}(t)$. Using their conclusion in our case, we have
\begin{equation}
\Sigma(t)\sim |\nu(t)| \ 
\label{eq5.2.7}
\end{equation}
for fat-tailed traded volumes. Thus, the correlation of the volume imbalance in equation~(\ref{eq5.2.2}) is approximately 
\begin{equation} 
\mathrm{corr}\big(\nu_k(t),\nu_l(t)\big)\sim\big\langle P_{kl}(t)\Sigma_k(t)\Sigma_l(t)\big\rangle_t  \ .
\label{eq5.2.8}
\end{equation}
\begin{figure*}[tb]
\begin{center}
\includegraphics[width=0.6\textwidth]{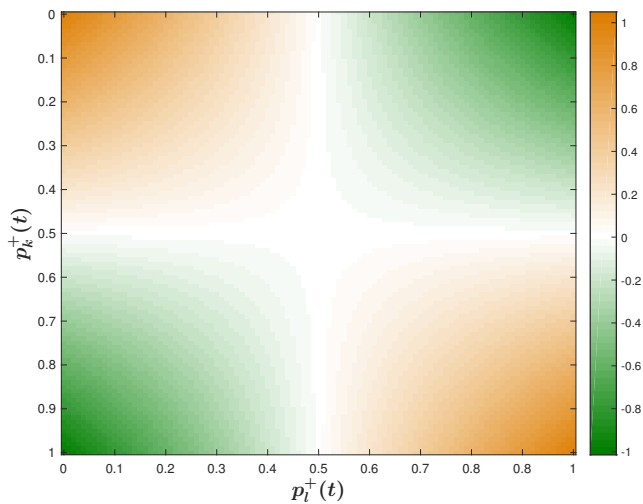}
\vspace*{-0.3cm}
\caption{The contour of $P_{kl}(t)$ depending on $p_k^+(t)$ and $p_l^+(t)$. The value of $P_{kl}(t)$ is indicated by the color.}
\vspace*{-0.5cm}
\label{fig54}
\end{center}
\end{figure*}
We analyze the dependence of $P_{kl}(t)$ on $p_k^+(t)$ and $p_l^+(t)$, see figure~\ref{fig54}. For very small volume imbalances, the probability of a surplus of volumes bought is very close to the one of a surplus of volumes sold in time $t$, \textit{i.e.}  $p_k^+(t)\approx p_k^-(t)\approx 0.5$. For this case, $P_{kl}(t)$ tends to zero, as seen in figure~\ref{fig54}. Accordingly, the correlation of volume imbalances goes towards zero according to equation~(\ref{eq5.2.8}). A correlation coefficient around zero indicates that the positive dependencies on the copulas are comparable to the negative ones as shown in figure~\ref{fig53}. Consequently, the very small local fluctuations, positively correlated with the absolute values of the very small volume imbalances, result in a similar strength of dependencies in the four corners of the copula density, see figure~\ref{fig51}. In contrast, the very large volume imbalances, corresponding to the very large local fluctuations according to equation~(\ref{eq5.2.7}), imply a high probability for most of the traded volumes being bought or sold. When both stocks $k$ and $l$ have very large volume imbalances, we find a rather high effective weight $P_{kl}(t)$ at the four corners of figure~\ref{fig54}. As a result, the very large local fluctuations in both stocks together with a high value of $P_{kl}(t)$ lead to a considerable correlation of volume imbalances. This correlation turns out to be positive, as the positive dependencies prevailing over negative ones result in a positive asymmetry $\gamma_{kl}$, corresponding to a positive correlation in figure~\ref{fig53}.

\begin{figure*}[p]
\raggedleft
\includegraphics[width=0.85\textwidth]{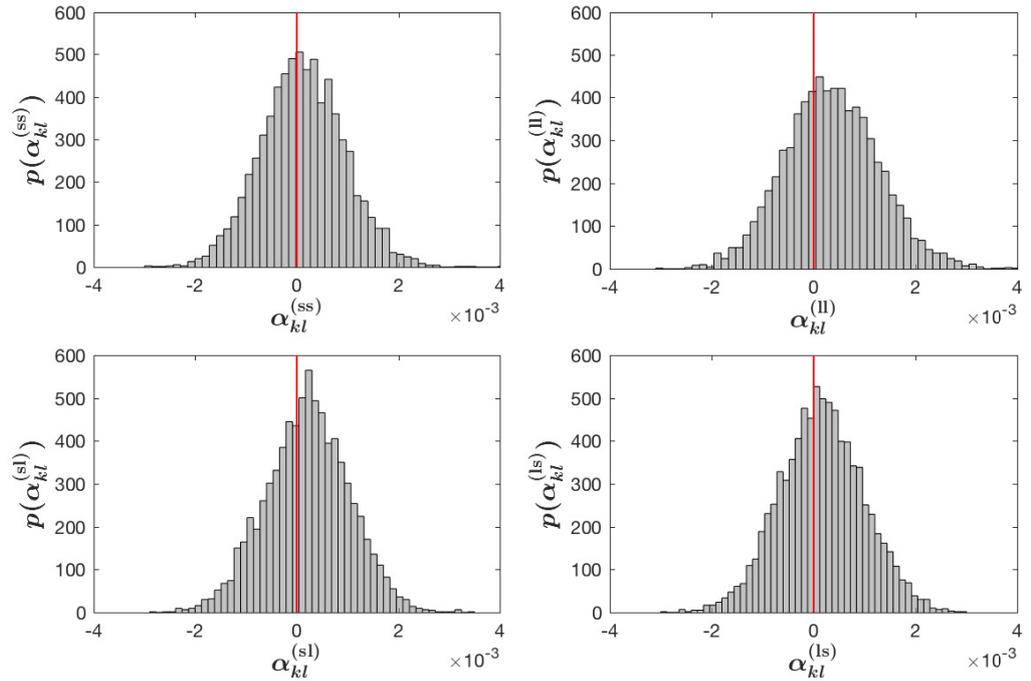}
\caption{Histograms of asymmetry values of 4950 stock pairs ($k$, $l$) for positive dependence $p(\alpha_{kl}^{(\mathrm{ss})})$,  $p(\alpha_{kl}^{(\mathrm{ll})})$,  $p(\alpha_{kl}^{(\mathrm{sl})})$, and $p(\alpha_{kl}^{(\mathrm{ls})})$, corresponding to the copula densities $\mathrm{cop}^{(\mathrm{ss})}(q_1,q_2)$, $\mathrm{cop}^{(\mathrm{ll})}(q_1,q_2)$, $\mathrm{cop}^{(\mathrm{sl})}(q_1,q_2)$, and $\mathrm{cop}^{(\mathrm{ls})}(q_1,q_2)$, respectively. All the histograms are normalized to one.}
\label{fig55}
\end{figure*}
\begin{figure*}[htbp]
\raggedleft
\includegraphics[width=0.85\textwidth]{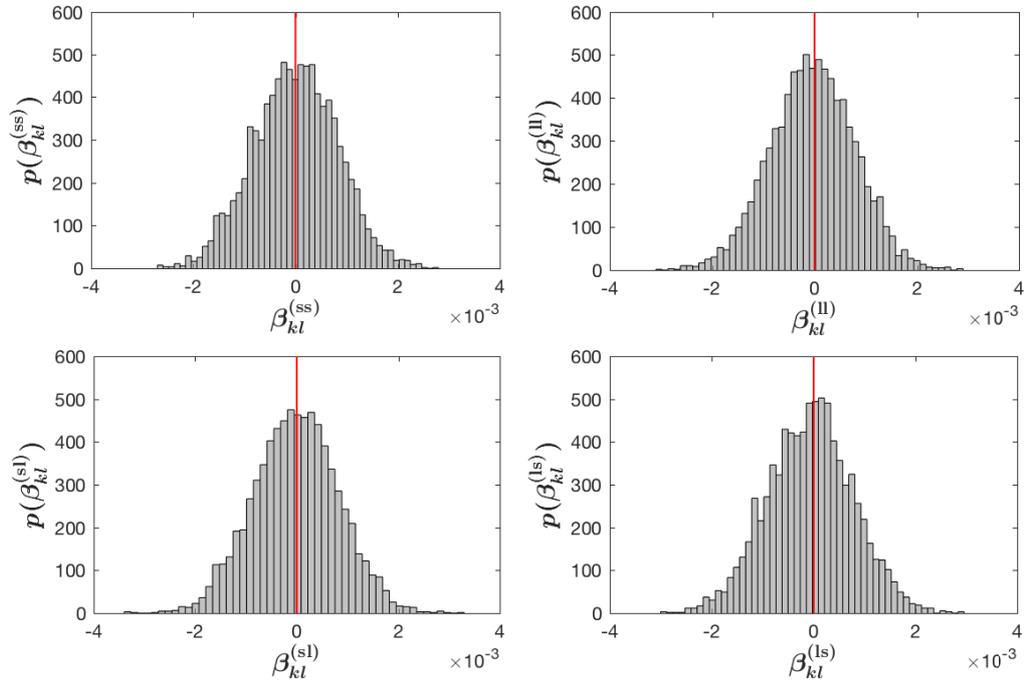}
\caption{Histograms of asymmetry values of 4950 stock pairs ($k$, $l$) for negative dependence $p(\beta_{kl}^{(\mathrm{ss})})$,  $p(\beta_{kl}^{(\mathrm{ll})})$,  $p(\beta_{kl}^{(\mathrm{sl})})$, and $p(\beta_{kl}^{(\mathrm{ls})})$, corresponding to the copula densities $\mathrm{cop}^{(\mathrm{ss})}(q_1,q_2)$, $\mathrm{cop}^{(\mathrm{ll})}(q_1,q_2)$, $\mathrm{cop}^{(\mathrm{sl})}(q_1,q_2)$, and $\mathrm{cop}^{(\mathrm{ls})}(q_1,q_2)$, respectively. All the histograms are normalized to one.}
\label{fig56}
\end{figure*}

\subsection{Influence on the asymmetries of tail dependencies}
\label{sec54}

In section~\ref{sec3}, we quantified and analyzed the asymmetries of tail dependencies between stocks. Here, we want to find out how the large local fluctuations act on the tail asymmetries in the copula density, characterized by $\alpha_{kl}$ and $\beta_{kl}$ for positive and negative dependencies, respectively. We work out the distributions of $\alpha_{kl}$ and $\beta_{kl}$ for four conditional copula densities, defined in equation~(\ref{eq5.1.1}), and show the results in figures~\ref{fig55} and \ref{fig56}. For the negative dependencies in figure~\ref{fig56}, the overall asymmetries are not pronounced in the four distributions $p(\beta_{kl}^{(\mathrm{ss})})$, $p(\beta_{kl}^{(\mathrm{ll})})$,  $p(\beta_{kl}^{(\mathrm{sl})})$,  $p(\beta_{kl}^{(\mathrm{ls})})$. Their skewness in table~\ref{tab1} is relatively small and changes slightly compared to the skewness of the distributions of $\alpha_{kl}$. For the positive dependencies, the overall asymmetries depend on the local fluctuations, see figure~\ref{fig55}. If both stocks of a pair have small local fluctuations, the skewness of the distribution $p(\alpha_{kl}^{(\mathrm{ss})})$ is 0.0665, which is smaller than the value of 0.0977 in the unconditional copula density, defined in equation~(\ref{eq3.2.2}). If both stocks have large local fluctuations, an overall right shift of the distribution $p(\alpha_{kl}^{(\mathrm{ll})})$ shows up with a skewness of 0.1247. If one stock has large local fluctuations and the other one has small local fluctuations, independently of the symmetry or asymmetry of $p(\beta_{kl}^{(\mathrm{sl})})$ and $p(\beta_{kl}^{(\mathrm{ls})})$, we find a very small skewness for them. Among the four distributions for positive dependence, only $p(\alpha_{kl}^{(\mathrm{ll})})$ exhibits sizeable right shift and a positive fat tail, implying that the large local fluctuations in both stocks contribute to the dependence of demands more than the one of supplies, probably indicating a bull market.

\section{Conclusions}
\label{sec6}

We investigated the influence of large local fluctuations on the dependence of demands between stocks. The demand is quantified by the volume imbalance, where the positive demand is due to a surplus of volumes bought, while the negative demand is the supply if a surplus of the volumes is sold. We employed copulas to study the dependence of demands, and found stronger positive dependencies than negative ones. Hence, if the demand for one stock is large, it is likely to find large demand for other stocks as well. The situation is analogous for supplies. The bivariate $\mathcal{K}$ copula density function describes the empirical copula density better than the Gaussian one, especially the fat-tailed dependencies. The bivariate $\mathcal{K}$ copula density function follows from a random matrix model and only depends on two parameters, an average correlation coefficient $c$ and a parameter $N$ measuring the strength of the fluctuations of the correlations.

We discussed the empirical copula densities conditioned on the local noise intensities, and found that the extremely large local fluctuations from both stocks of a pair strengthen the positive dependencies of demands but weaken the negative ones. We attributed this interesting feature to the cross-correlation of volume imbalances between stocks, which in turn is related to large local fluctuations and signs of the volume imbalances. We uncover that the larger the local fluctuations, the stronger is the cross-correlation of volume imbalances, and the bigger is the difference between positive and negative dependencies of demands in the copula densities.

We also looked at the asymmetries of tail dependencies of demands. They are not pronounced for negative dependencies but sizeable for the positive ones. For the latter, the large local fluctuations cause a shift from zero to the right in the distribution of the asymmetries. We therefore conclude that large local fluctuations influence the dependence of demands more than the dependence of supplies, probably reflecting a bull market with persistent increase of prices in the markets.

\section*{Acknowledgements}
\addcontentsline{toc}{section}{Acknowledgements}

One of us (SW) acknowledges financial support from the China Scholarship Council (grant no. 20130689\-0014).

\appendix
\addtocontents{toc}{\protect\setcounter{tocdepth}{0}}
\section{Stock information}
\addtocontents{toc}{\protect\setcounter{tocdepth}{1}}
\addcontentsline{toc}{section}{Appendix A. Stock information}

\label{appA}

With the TAQ data set, we calculate the average number of daily trades for 496 available stocks from S$\&$P 500 index in 2008. The first 100 stocks with the largest average number of daily trades are listed in table~\ref{tabA}, where records in detail the information of symbols, economic sectors and the average numbers of daily trades for each stock.

\begin{footnotesize}

\setlength{\tabcolsep}{9.2pt}
\begin{longtable}[p]{llllll}
\caption{\label{tabA}The first 100 stocks with the largest average number of daily trades} \\
\hlineB{2}
Stocks& Sectors & Numbers& Stocks& Sectors & Numbers\\
\hline
\endfirsthead
\caption* {Table \ref{tabA}: (continued)}\\
\hlineB{2}
Stocks& Sectors & Numbers& Stocks& Sectors & Numbers\\
\hline
\endhead
\hlineB{2}
\endfoot
\endlastfoot
%Stocks& Sectors & Numbers& Stocks& Sectors & Numbers\\ %[0.5ex]
%\hline
C	&	Financials	&	98990.8	&	AMGN	&	HealthCare	&	21715.6	\\
BAC	&	Financials	&	90648.7	&	SPLS	&	ConsumerDiscretionary	&	21528.3	\\
AAPL&	InformationTechnology	&	86242.5	&	SBUX	&	ConsumerDiscretionary	&	21420.5	\\
MSFT&	InformationTechnology	&	80399.8	&	GILD	&	HealthCare	&	20880.6	\\
JPM	&	Financials	&	75825.8	&	FCX	&	Materials	&	20762.8	\\
WFC	&	Financials	&	68118.3	&	SYMC	&	InformationTechnology	&	20490.1	\\
INTC	&	InformationTechnology	&	63849.0	&	NCC	&	Financials	&	20029.7	\\
GE	&	Industrials	&	61435.8	&	GLW	&	InformationTechnology	&	19865.2	\\
CSCO	&	InformationTechnology	&	60952.6	&	DIS	&	ConsumerDiscretionary	&	19754.3	\\
WB	&	Financials	&	60803.0	&	ADBE	&	InformationTechnology	&	19180.9	\\
XOM	&	Energy	&	56978.8	&	TGT	&	ConsumerDiscretionary	&	19068.0	\\
MER	&	Financials	&	55616.2	&	KO	&	ConsumerStaples	&	18812.8	\\
AIG	&	Financials	&	48129.0	&	VLO	&	Energy	&	18770.7	\\
QCOM	&	InformationTechnology	&	47234.5	&	F	&	ConsumerDiscretionary	&	18741.8	\\
ORCL	&	InformationTechnology	&	45197.8	&	SLB	&	Energy	&	18712.4	\\
MS	&	Financials	&	42930.3	&	SNDK	&	InformationTechnology	&	18464.7	\\
YHOO&	InformationTechnology	&	39279.2	&	ALTR	&	InformationTechnology	&	18444.2	\\
WMT&	ConsumerStaples	&	37852.5	&	XLNX	&	InformationTechnology	&	18187.2	\\
DELL&	InformationTechnology	&	36807.0	&	BMY	&	HealthCare	&	17949.7	\\
T	&	TelecommunicationsServices	&	36013.4	&	SGP	&	HealthCare	&	17947.4	\\
CMCSA&	ConsumerDiscretionary	&	35446.6	&	DTV	&	ConsumerDiscretionary	&	17933.3	\\
PFE	&	HealthCare	&	31997.7	&	RF	&	Financials	&	17566.6	\\
NVDA&	InformationTechnology	&	31618.3	&	MOT	&	InformationTechnology	&	17293.7	\\
AMAT&	InformationTechnology	&	31156.6	&	HCBK	&	Financials	&	17177.7	\\
HD	&	ConsumerDiscretionary	&	30661.2	&	NTAP	&	InformationTechnology	&	17017.8	\\
HAL	&	Energy	&	30160.6	&	XTO	&	Energy	&	16897.1	\\
HPQ	&	InformationTechnology	&	29049.2	&	GOOG	&	InformationTechnology	&	16870.1	\\
CHK	&	Energy	&	28869.6	&	MO	&	ConsumerStaples	&	16818.3	\\
USB	&	Financials	&	28501.7	&	CVS	&	ConsumerStaples	&	16129.3	\\
BRCM	&	InformationTechnology	&	28333.2	&	JAVA	&	InformationTechnology	&	15962.9	\\
CVX	&	Energy	&	28211.5	&	BBBY	&	ConsumerDiscretionary	&	15786.4	\\
EMC	&	InformationTechnology	&	27682.0	&	BK	&	Financials	&	15589.9	\\
EBAY&	InformationTechnology	&	27589.1	&	LLTC	&	InformationTechnology	&	15450.5	\\
SCHW&	Financials	&	25703.3	&	WFT	&	Energy	&	15316.3	\\
AA	&	Materials	&	25148.4	&	MU	&	InformationTechnology	&	14973.1	\\
TXN	&	InformationTechnology	&	24315.1	&	HBAN	&	Financials	&	14899.5	\\
GS	&	Financials	&	24113.1	&	MCD	&	ConsumerDiscretionary	&	14896.6	\\
COP	&	Energy	&	24010.7	&	COST	&	ConsumerStaples	&	14812.1	\\
PG	&	ConsumerStaples	&	23998.2	&	UNH	&	HealthCare	&	14685.2	\\
VZ	&	TelecommunicationsServices	&	23540.2	&	DOW	&	Materials	&	14684.0	\\
AXP	&	Financials	&	23508.4	&	NBR	&	Energy	&	14642.2	\\
AMZN	&	ConsumerDiscretionary	&	23213.6	&	COF	&	Financials	&	14577.6	\\
FITB	&	Financials	&	23105.8	&	KFT	&	ConsumerStaples	&	14542.3	\\
JNPR	&	InformationTechnology	&	22952.1	&	AMD	&	InformationTechnology	&	14516.8	\\
GM	&	ConsumerDiscretionary	&	22379.9	&	GPS	&	ConsumerDiscretionary	&	14501.0	\\
TWX	&	ConsumerDiscretionary	&	22075.0	&	OXY	&	Energy	&	14166.3	\\
LOW	&	ConsumerDiscretionary	&	21933.8	&	CAT	&	Industrials	&	14003.3	\\
JNJ	&	HealthCare	&	21906.1	&	M	&	ConsumerDiscretionary	&	13884.9	\\
MRK	&	HealthCare	&	21903.3	&	DD	&	Materials	&	13859.4	\\\hline
S	&	TelecommunicationsServices	&	21724.9	&	DHI	&	ConsumerDiscretionary	&	13810.6	\\
\hlineB{2}
\end{longtable}
\end{footnotesize}
%\normalsize

\end{document}